\documentclass[preprint,showpacs,showkeys,preprintnumbers,prb,fleqn,superscriptaddress]{revtex4-2}

\usepackage{hyperref}
\usepackage{tabularx}
\usepackage[dvipsnames]{xcolor}
\usepackage[normalem]{ulem}
\usepackage{graphicx}
\usepackage{dcolumn}
\usepackage{amsmath}
\usepackage[english]{babel}
\usepackage[utf8]{inputenc}
\usepackage{stmaryrd}
\usepackage{mathtools}
\DeclarePairedDelimiter\bra{\langle}{\rvert}
\DeclarePairedDelimiter\ket{\lvert}{\rangle}

\begin{document}

\title{Magnetic phases and zone-folded phonons in a frustrated van der Waals magnet}

\author{Amit Pawbake}
\thanks{These two authors contributed equally}
\affiliation{LNCMI, UPR 3228, CNRS, EMFL, Universit\'e Grenoble Alpes, 38000 Grenoble, France}

\author{Florian Petot}
\thanks{These two authors contributed equally}
\affiliation{Aix-Marseille Universit\'e, CNRS, CINaM, Marseille, France}

\author{Florian Le Mardel\'e}
\affiliation{LNCMI, UPR 3228, CNRS, EMFL, Universit\'e Grenoble Alpes, 38000 Grenoble, France}

\author{Tristan Riccardi}
\affiliation{LNCMI, UPR 3228, CNRS, EMFL, Universit\'e Grenoble Alpes, 38000 Grenoble, France}
\affiliation{Universit\'e Grenoble Alpes, CNRS, Grenoble INP, Institut NÉEL, 38000 Grenoble, France}

\author{Julien L\'ev\^eque}
\affiliation{Aix-Marseille Universit\'e, CNRS, CINaM, Marseille, France}

\author{Benjamin A. Piot}
\affiliation{LNCMI, UPR 3228, CNRS, EMFL, Universit\'e Grenoble Alpes, 38000 Grenoble, France}

\author{Milan Orlita}
\affiliation{LNCMI, UPR 3228, CNRS, EMFL, Universit\'e Grenoble Alpes, 38000 Grenoble, France}

\author{Johann Coraux}
\affiliation{Universit\'e Grenoble Alpes, CNRS, Grenoble INP, Institut NÉEL, 38000 Grenoble, France}

\author{Michal Hubert}
\affiliation{Faculty of Mathematics and Physics, Institute of Physics, Charles University, Ke Karlovu 5, 121 16 Prague 2, Czech Republic}

\author{Jan Dzian}
\affiliation{LNCMI, UPR 3228, CNRS, EMFL, Universit\'e Grenoble Alpes, 38000 Grenoble, France}
\affiliation{Faculty of Mathematics and Physics, Institute of Physics, Charles University, Ke Karlovu 5, 121 16 Prague 2, Czech Republic}

\author{Martin Veis}
\affiliation{Faculty of Mathematics and Physics, Institute of Physics, Charles University, Ke Karlovu 5, 121 16 Prague 2, Czech Republic}

\author{Yurii Skourski}
\affiliation{Hochfeld-Magnetlabor Dresden (HLD-EMFL), Helmholtz-Zentrum Dresden-Rossendorf, 01328 Dresden, Germany}

\author{Bing Wu}
\affiliation{Department of Inorganic Chemistry, University of Chemistry and Technology Prague, Technicka 5, 166 28 Prague 6, Czech Republic}

\author{Zdenek Sofer}
\affiliation{Department of Inorganic Chemistry, University of Chemistry and Technology Prague, Technicka 5, 166 28 Prague 6, Czech Republic}

\author{Beno\^it Gr\'emaud}
\affiliation{Aix Marseille Univ, Universit\'e de Toulon, CNRS, CPT, Marseille, France}

\author{Andr\'es Sa\'ul}
\email{andres.saul@cnrs.fr}
\affiliation{Aix-Marseille Universit\'e, CNRS, CINaM, Marseille, France}

\author{Cl\'ement Faugeras}
\email{clement.faugeras@lncmi.cnrs.fr}
\affiliation{LNCMI, UPR 3228, CNRS, EMFL, Universit\'e Grenoble Alpes, 38000 Grenoble, France}

\date{\today }

\begin{abstract}
2D magnetic materials have attracted extensive research interest due to their potential application in nanospintronics, optospintronics, and in magnonics. Ferromagnetic as well as antiferromagnetic layered materials have been demonstrated and successfully inserted into van der Waals heterostructures. However, the effects of magnetic frustration in van der Waals materials and the possibilities offered by spin configurations characterized by nonlinear spin arrangements have not been fully considered yet. Herein, we establish the magnetic phase diagram of bulk CrOCl, a frustrated van der Waals magnet, using magnetization and magneto-optical spectroscopy techniques. In particular, we use the magnetic superstructures relative to the crystallographic unit cell and the associated rich zone-folded phonon series to describe the magnetic field induced phases. Theoretical calculations taking into account the competing nearest neighbors magnetic exchange interactions provide a unique insight into the lattice vibrations of this class of magnetic system. This study expands the scope of 2D magnetic materials and provides a methodology to characterize frustrated van der Waals magnets.
\end{abstract}

\keywords{van der Waals magnet, Raman scattering, zone folded phonons, Density functional theory, magnetic phases}

\maketitle

\section{Introduction}
\label{sec:intro}

On-lattice frustration, in certain magnetic compounds featuring spin-spin interactions of various kinds between first, second, etc, neighbor lattice sites, leads to very rich magnetic phase diagrams comprising highly-degenerate disordered phases and complex spin orders. Varying the temperature, magnetic field or hydrostatic pressure, one can navigate through these exotic phase diagrams. Spectacular examples of emerging exotic quantum phases~\cite{Zapf2014,Giamarchi2008} and of cascades of magnetic ground states driven by an external magnetic field~\cite{Shiramura1998,Mitamura2012,Wosnitza2016,Wulferding2020,Sahasrabudhe2020}, or hydrostatic pressure~\cite{Thede2014,Schaller2023} have been reported in the last years.

Van der Waals magnetic materials are layered materials that can be thinned down to the monolayer limit, while retaining collective magnetic properties~\cite{Huang2017}, different in nature from those of the bulk parent material~\cite{Fei2018}. They altogether offer a broad portfolio of intralayer magnetic ordered states such as ferromagnetism (FM), Néel-like zig-zag or stripe antiferromagnetism~\cite{Huang2017,Kurosawa1983,Leflem1982} (AFM) combined with ferro- or antiferromagnetic interlayer interactions~\cite{Jiang2021}. Accordingly, they represent a rich platform to revisit and further explore low-dimensional magnetism, with new opportunities as they can be inserted within van der Waals heterostructures~\cite{Geim2013} to induce specific properties in the neighboring layers via proximity effects~\cite{Zhong2017,Ciorciaro2020,Huang2020}; they may even pave the way to an ultimately thin (van der Waals) spintronics~\cite{Sierra2021}.

Depending on the lattice structure and on the nature/strength of spin-spin interactions between different kinds of neighbor sites, these interactions may have antagonist effects, each promoting alone a different kind of order. Such competition of interactions prevents simple (i.e. ferro or antiferromagnetic) spin orders to develop, and instead favors nontrivial spin arrangements. Bulk chromium oxychloride CrOCl falls in this class of compounds -- it is a frustrated magnet, possessing a rich phase diagram whose phases have only been partly assigned.

\begin{figure*}[ht!]
\includegraphics[width=1\linewidth,angle=0,clip]{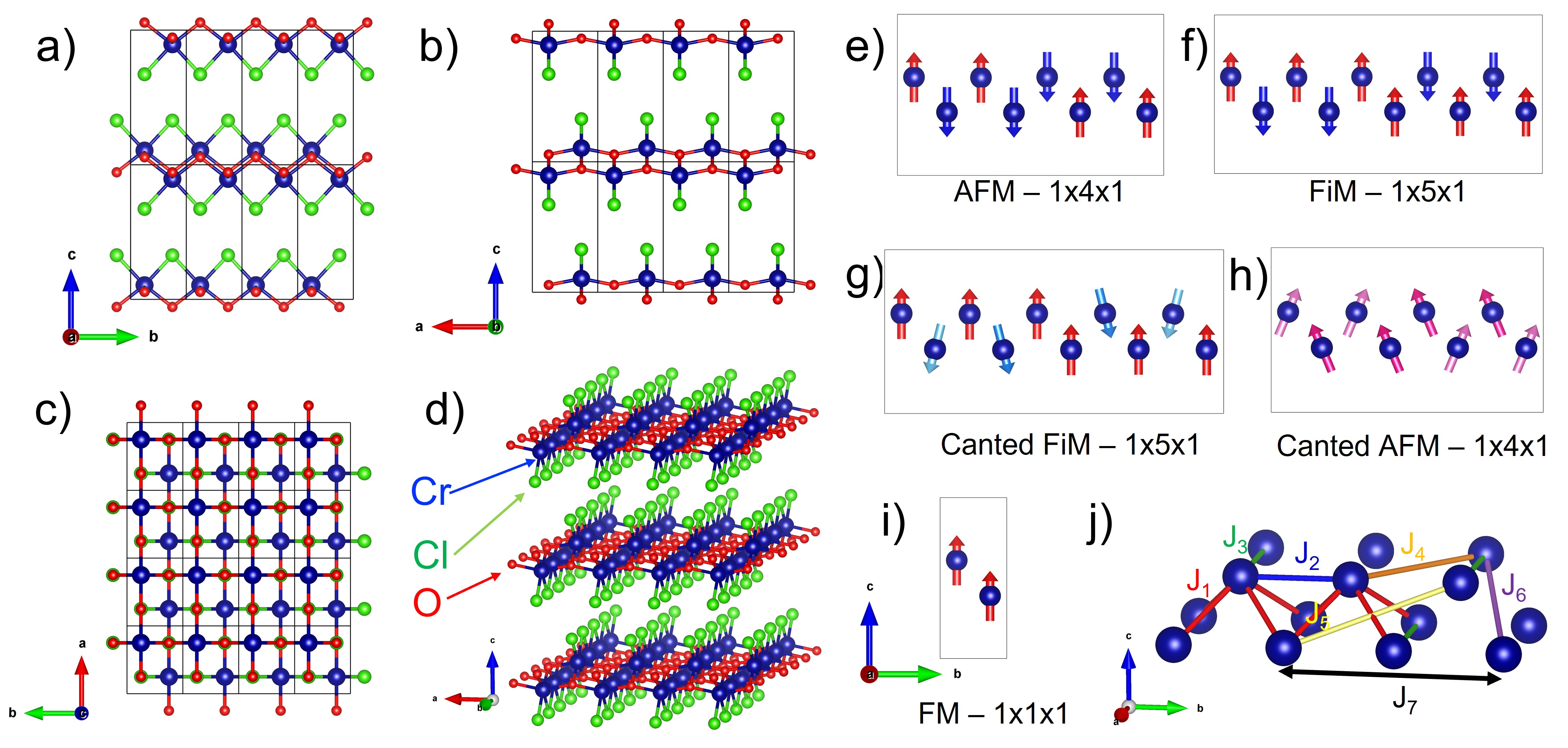}
\caption{a-c) Crystal structure of bulk CrOCl along the main crystallographic axes and d) perspective view of the lattice. The field-induced magnetic phases of CrOCl considered in this work:
e) Antiferromagnetic (AFM) with a 4$\vec{b}$ period;
f) Ferrimagnetic (FiM) with a 5$\vec{b}$ period compatible with a 1/5 total magnetization;
g) canted FiM phase with a 5$\vec{b}$ period;
h) canted AFM phase with a 4$\vec{b}$ period;
i) Ferromagnetic (FM);
j) Definition of the Cr-Cr magnetic exchange parameters $J_i$.
\label{fig:Struct-and-order}}
\end{figure*}

At room temperature, CrOCl crystallizes in the orthorhombic FeOCl structure with \textit{Pmmn} space group~\cite{Forsberg1962,Christensen1975} (see Table S1 of Supporting information). The crystal structure of CrOCl together with the relevant views along the  $\vec{a}$, $\vec{b}$, and $\vec{c}$ axis are presented in Fig.~\ref{fig:Struct-and-order}a-d. When decreasing temperature, magnetic correlations become significant below $27$~K where an incommensurate spin density wave builds up~\cite{Patoche2014}, which then evolves below the N\'{e}el temperature $T_\mathrm{N}=14$~K into an antiferromagnetic (AFM) state with a 4$\vec{b}$ period~\cite{Christensen1975} (see Fig.~\ref{fig:Struct-and-order}e). Even though the size of the magnetic unit cell is 4$\vec{b}$, X-ray experiments detected a 2$\vec{b}$ structural distortion~\cite{Angelkort2009} accompanied by a distortion to a monoclinic lattice with \textit{P2$_1$/m} space group.

The 4$\vec{b}$ periodicity of the magnetic order is a consequence of the frustration in the magnetic exchange parameters as defined in Fig.~\ref{fig:Struct-and-order}j. Exchange interactions calculated using density functional theory methods and the broken symmetry formalism \cite{Radtke2010a,Saul2011,Okada2016,Saul2018,Vaclavkova2020,Rocquefelte2023}, i.e., by mapping total energies corresponding to various collinear spin arrangements within a supercell onto the Heisenberg Hamiltonian, are presented in the S3 of the Supporting Information. The leading exchange interactions are $J_1$, $J_2$, $J_3$, and $J_7$. As can be seen in Fig.~\ref{fig:Struct-and-order}j, the FM interaction $J_3$ connects Cr atoms in different unit cells along the $a$ axis and does not lead to a new periodicity along this direction. Frustration is due to incompatible interactions $J_1$ (AFM), $J_2$ (FM), and $J_7$ (AFM). In particular $J_7$ is the main responsible for stabilizing the four times periodicity along the $\vec{b}$ axis.

When a magnetic field is applied along the $\vec{c}$ axis, bulk CrOCl undergoes a series of magnetic phase transitions first to a spin-flop phase around $B=3.0$~T and then to a ferrimagnetic (FiM) phase (see Fig. \ref{fig:Struct-and-order}f) above $B=3.85$~T. This phase has been inferred from the observation of a plateau in the magnetization~\cite{Gu2022} corresponding to $1/5$ of the magnetization at saturation. For $B>$~4~T, an apparent splitting of a phonon with $A_g$ symmetry~\cite{Gu2022} has been assigned to a Raman scattering signature of the FiM phase. Above the FiM state, the high magnetic field phases of CrOCl are not known even though it has been shown that the magnetic lattice saturates at $B=30$~T~\cite{Patoche2014}. Recently, thin layers of CrOCl down to the monolayer have been investigated by magneto-tunneling measurements~\cite{Zhang2023} suggesting magnetocrystalline anisotropies more subtle than the easy-axis anisotropy for bulk CrOCl~\cite{Das2022,Zhang2023}.

In this work, we present magneto-Raman scattering and magneto-infrared (IR) absorption spectroscopy analysis of phonon modes in bulk CrOCl, confronted to magnetization-versus-magnetic field measurements up to high magnetic fields (30~T continuous, 60~T pulsed) and density functional theory (DFT) calculations. Our results show that the magnetic frustration in this van der Waals magnet induces magnetic phases with unit cells much larger than the crystallographic one. This additional magnetic periodicity changes both the phonon Raman scattering and the infrared responses by folding the phonon Brillouin zone onto the $\Gamma$ point. We observe a series of zone-folded phonons (ZFP) in the low temperature AFM phase, which changes under an external magnetic field into another series of ZFP modes in the FiM phase and which persists up to $B=19$~T. Above this magnetic field, the series of ZFP modes transforms again into another series of ZFP similar to that observed at $B=0$ indicating, at high magnetic fields, the presence of a magnetic phase with the same periodicity as the AFM phase. For $B>19$~T, all Raman active phonon experience a gradual change of their energies, as large as $1$~meV, up to the saturation magnetic field of $B=30$~T. In particular, the ZFP modes allow for a spectroscopy of the magnetic ground states and also reflect the magnetic hysteresis as observed when sweeping the magnetic field in opposite directions.

Using DFT calculations, we describe the phonon modes of these different ground states. The theoretical calculations of the ZFP modes compare well with the experimental observations supporting the presence of magnetic phases with unit cells much larger than the crystallographic one. Using an \textit{unfolding} procedure~\cite{Allen2013} we were able to assign the series of ZFP to the $\vec{k} \neq \vec{0}$  phonon modes of the crystallographic cell.

Finally, experiments performed with the external magnetic field applied along the in-plane directions show that bulk CrOCl has an easy magnetic axis along $\vec{c}$, but also an intermediate and a hard axis in the plane of the layers.

\begin{figure*}[ht!]
\includegraphics[width=0.9\linewidth,angle=0,clip]{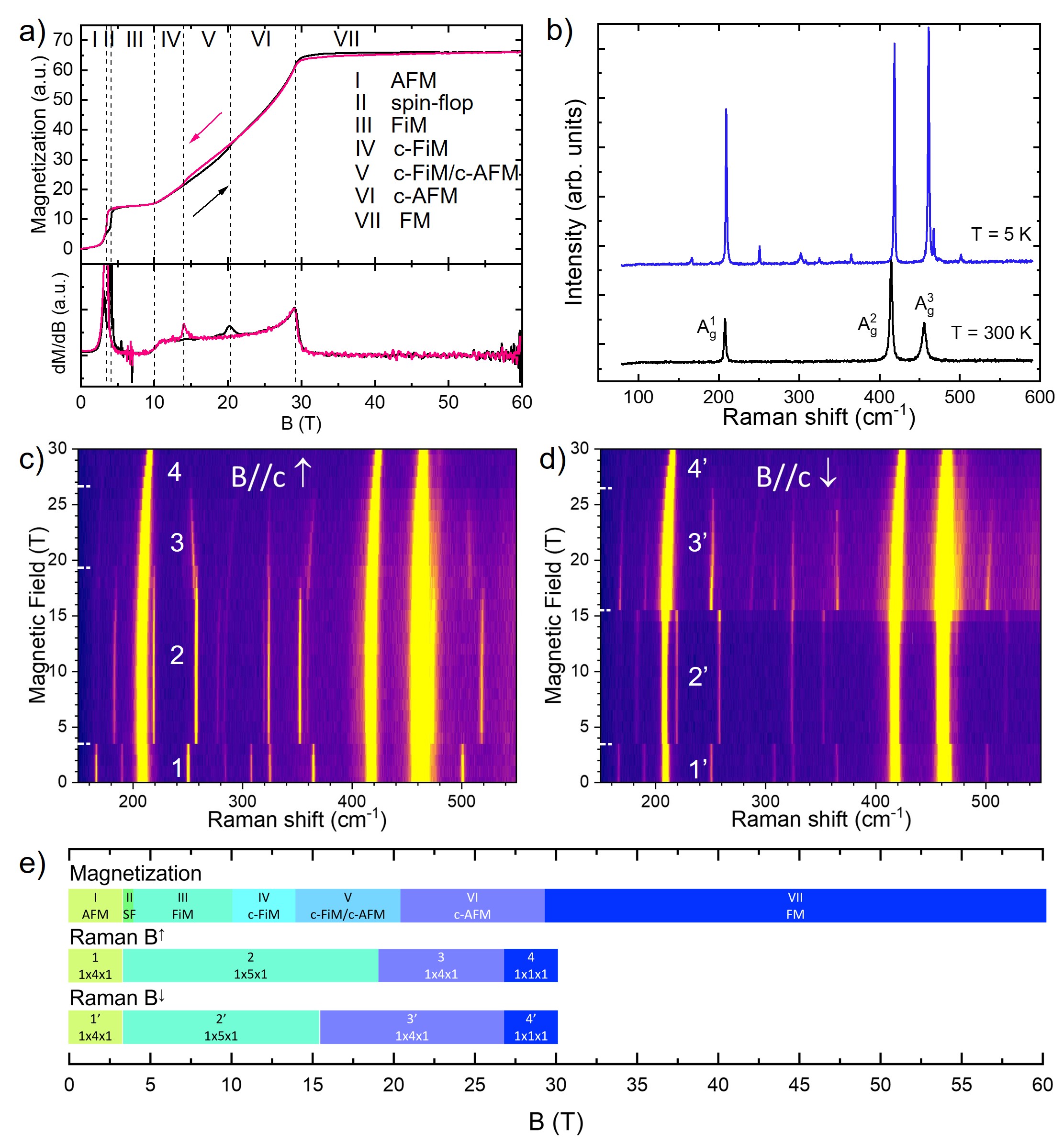}
\caption{a) Low temperature ($4.5$~K) magnetization curve of bulk CrOCl with the magnetic field applied along $\vec{c}$. The data for $B<7$~T have been measured with SQUID magnetometry, and in pulsed magnetic fields for $B>7$~T, see Fig.~S8 of Supporting Information. Vertical dashed lines indicate the critical magnetic fields separating the different magnetic regions labeled by roman numbers. b) Raman scattering response of bulk CrOCl measured at room temperature and at $T=5$~K. c-d) Magneto-Raman scattering spectra acquired when sweeping the magnetic field along $\vec{c}$ (c) from $0$ to $30$~T and (d) from $30$~T to $0$ measured with 1~T field increments/decrements. Four different regions can be defined when sweeping the magnetic field up or down, labeled with arabic and primed arabic labels respectively (see text). e) Schematic comparison of the different regions associated to specific behaviors in the magnetization versus magnetic field measurements and the Raman scattering response when sweeping the magnetic field up and down as shown in panels a, c and d.
\label{fig:MvsB-and-Raman}}
\end{figure*}

\section{Magnetization measurements}
\label{sec:m-vs-b}

Figure~\ref{fig:MvsB-and-Raman}a shows the magnetization (M) as a function of a magnetic field applied along the $\vec{c}$ axis, measured up to $B = 60$~T.
Seven different regions can be defined in the $M$ vs $B$ plot. They have been labeled with roman numbers in Fig.~\ref{fig:MvsB-and-Raman}a.
As mentioned above, the zero field phase, region I in Fig.~\ref{fig:MvsB-and-Raman}a, is an AFM phase (see Fig.\ref{fig:Struct-and-order}e) with a 4$\vec{b}$ period \cite{Christensen1975}, which is maintained up to $B \sim 3.5$~T, in our experiment performed at $4.5$~K~\cite{Angelkort2009,Patoche2014,Zhang2019a,Gu2022,Zhang2023}.
Magnetization increases linearly as a function of B field in this phase. We understand this behavior as possibly originating from the finite temperature at which the experiment is performed, and to potential paramagnetic impurities in our bulk specimen.

A first hysteresis cycle is observed between $3.5$~T and $4$~T (region II in Fig.~\ref{fig:MvsB-and-Raman}a). When increasing the magnetic field in this region (black curve), the small plateau at around one tenth of the saturation magnetization has recently been assigned to the transition from the low temperature AFM state to a spin-flop phase, which then evolves to a FiM state characterized by a magnetization plateau (region III) at one fifth of the magnetization~\cite{Gu2022,Das2022,Xu2023,Zhang2023}.
%
The proposed magnetic order for the FiM phase in the $10$-spins unit cell is depicted in Fig.~\ref{fig:Struct-and-order}f, with $6$ spins (resp. $4$) aligned (resp. anti-aligned) with the field direction~\cite{Gu2022,Zhang2023}. It is interesting to note that this magnetic order with a 5$\vec{b}$ periodicity is the simplest conceivable one with a 1/5 of the total magnetization, but other orders might be relevant. The magnetization plateau associated with the ferrimagnetic phase extends up to $B=10$~T.

We have performed magneto-optical Kerr effect experiments that show the appearance of a net magnetization in CrOCl above $B=4$~T, see Fig.~S9 of Supporting information. The field dependent Kerr rotation spectra exhibit sudden increase in amplitude above $3.5$~T when a spectroscopic structure around $1.7$~eV appears. This corresponds to the Kerr rotation originating from interband transitions across the band gap in the ferrimagnetic phase. The ferrimagnetic origin of the polar Kerr rotation is proven by its sign change upon the change of the sign of external magnetic field and a relative saturation between $4.5$ and $6$~T.
Above $10$~T (regions IV, V, and VI in Fig.~\ref{fig:MvsB-and-Raman}a), the magnetic phases have not been identified so far. Our Raman scattering experiments and DFT calculations provides new insights on the magnetic order associated with them (see below). Around $B=29$~T magnetization saturates and remains constant (region VII) up to the highest value of the magnetic field accessible in our experiment, of $B=60$~T. CrOCl is then fully polarized in a FM state (see cartoon in Fig.~\ref{fig:Struct-and-order}i).

Evaluating the magnetic susceptibility dM/dB, see the lower panel of Fig.~\ref{fig:MvsB-and-Raman}a, we can identify two anomalies, at $B=20.5$~T when sweeping the magnetic field up, and at $B=14$~T when sweeping down. We understand these peaks in the susceptibility as signatures of magnetic phase transition which imply that, sweeping the magnetic field up, the canted FiM phase extends up to $B=20.5$~T, and that when sweeping the B field down, the canted AFM phase extends down to $B=14$~T. This magnetization curve will serve as a guide to interpret the results of Raman scattering and of infrared spectroscopy experiments.

\section{Raman scattering measurements}

\subsection{Temperature dependence}

The Raman scattering response of bulk CrOCl at $5$~K and $300$~K is presented in Fig.~\ref{fig:MvsB-and-Raman}b. At room temperature, it is composed of three contributions identified as phonons with $A_{g}$ symmetry~\cite{Zhang2019a} and labelled $A^{1-3}_g$. They correspond to atomic displacements along the $\vec{c}$ axis (see Fig.~S11 of Supporting Information). The other six Raman active modes with $B_{2g}$ and $B_{3g}$ symmetries and displacements along the $\vec{a}$ and $\vec{b}$ axis are expected to have extremely low intensities and cannot be observed in our experiment.

\begin{figure*}[ht!]
    \includegraphics[width=1.0\linewidth,angle=0,clip]{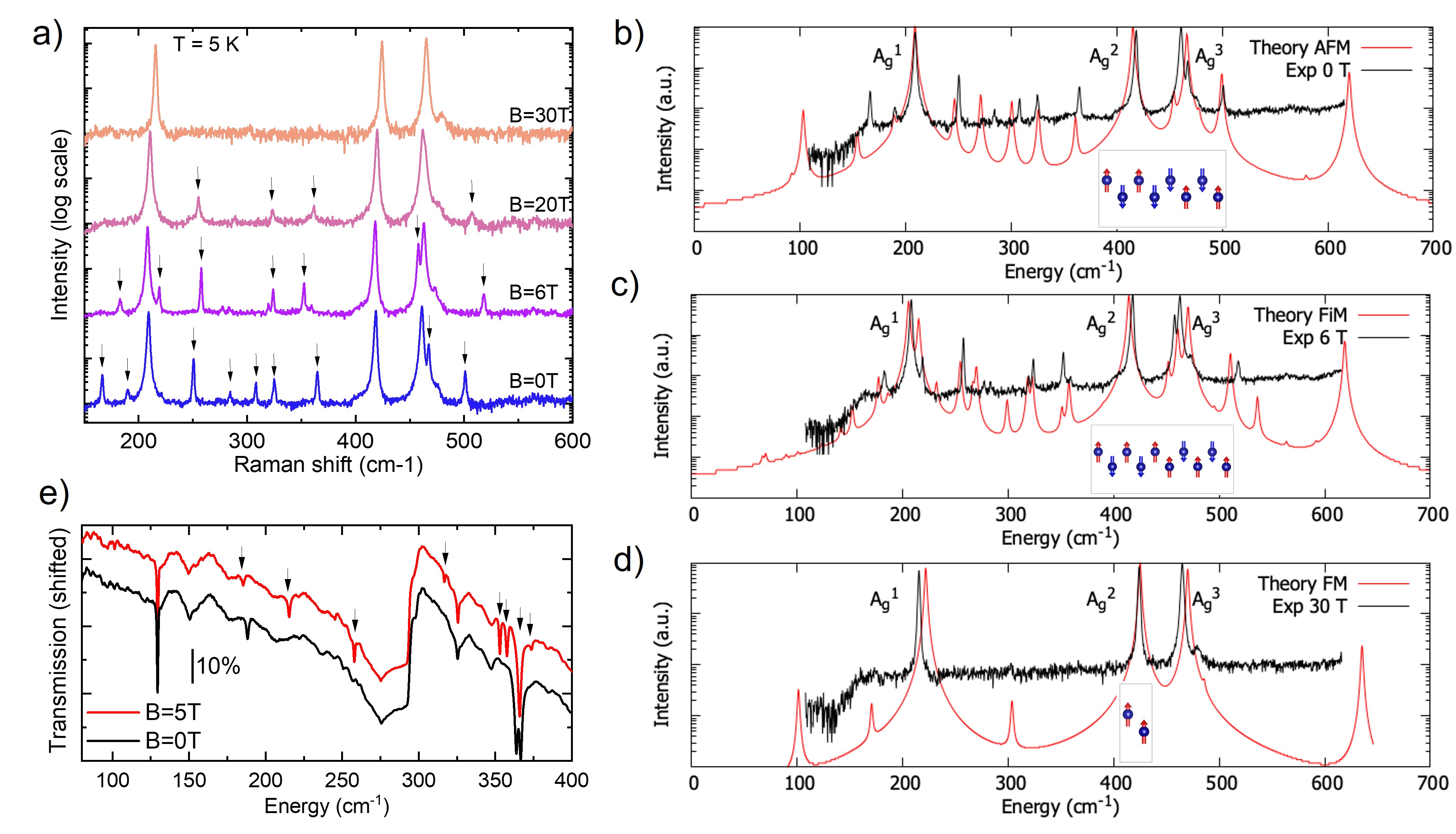}
    \caption{
        a) Low temperature Raman scattering response of bulk CrOCl at selected values of the magnetic field applied along $\vec{c}$ and corresponding to the AFM, to the FiM, the canted antiferromagnetic and to the FM phases (log scale). Arrows indicate the zone folded phonons.
        b-d) Experimental spectra (5~K) compared to theoretical calculations (log. scale; theoretical energies stretched by 6\%, structures used for calculation have been fully relaxed); b) measurement at 0~T, calculation for an AFM 1$\times$4$\times$1 monoclinic unit cell; c) measurement at 6~T, calculation for a FiM 1$\times$5$\times$1 orthorhombic unit cell; d) measurement at 30~T, calculation for a FM 1$\times$1$\times$1 orthorhombic unit cell; e) Low temperature IR transmission spectra measured at $B=0$~T (AFM phase) and at $B=5$~T (FiM phase). Arrows indicate the zone folded phonons.
    \label{fig:Raman-and-IR}}
\end{figure*}

When decreasing temperature, on top of these prominent phonon modes (hardened under the effect of anharmonicity), another series of Raman scattering features with much lower intensity can be seen clearly at $5$ K, see Fig.~\ref{fig:MvsB-and-Raman}b. The full temperature dependence can be found in Figs. S1 and S2 of the Supporting information. These modes become optically active when the long range magnetic order is settled and a magnetic unit cell, with a different size than the crystallographic cell, appears. This is the case when the AFM ground state with a 4$\vec{b}$ periodicity is stabilized below $T_N=14$~K. Figure~\ref{fig:Raman-and-IR}b) shows the comparison of the experimental Raman spectra at zero field and $5$ K and the DFT calculations for the fully relaxed 1$\times$4$\times$1 monoclinic unit cell with the AFM magnetic order shown in Fig.~\ref{fig:Struct-and-order}e. The calculated parameters of the relaxed AFM cell are presented in Table S2 of Supporting information. The calculation confirms that the low intensity Raman features arise from two effects: the change in symmetry, from orthorhombic to monoclinic, and the magnetoelastic distortions induced by the magnetic order, see Figs.~S9 and S10 of the Supporting information.

It is important to note, here, that the spin superperiodicity developed below $T_N$ does not explain, alone, the appearance of folded phonon modes in the Raman spectra because the electromagnetic wave does not interact with the spin but with the full charge density. As it is the case in CrOCl, most magnetic phase transitions are accompanied by small lattice distortions producing a superperiodicity (relative to the nonmagnetic crystal periodicity). The Raman activity of such modes depends on the amplitude of the lattice distortion, which is traced back to the strength of the magneto-elastic interaction. The additional features in the spectra thus reflect the different magnetic phases through the lattices distortion mentioned above, and can be used to reveal the field-induced magnetic phases.

\subsection{Magnetic field dependence}

The magnetic field dependence of the Raman scattering response for increasing fields is shown in Fig.~\ref{fig:MvsB-and-Raman}c, and in Fig.~\ref{fig:Raman-and-IR}a for 0, 6, 20 and 30~T.

When increasing the magnetic field, the series of additional features characteristic of the AFM-1$\times$4$\times$1 phase changes abruptly (within $100$~mT) to another series for $B > 3.85$~T  (see Fig.~S3 in Supporting information). This observation indicates that at this particular magnetic field, small displacements of the atomic positions occur following a change in the magnetic ground state and in the size of the magnetic unit cell. This is consistent with our understanding of the periodicity of the magnetic phases of bulk CrOCl that changes from 4$\vec{b}$ in the AFM phase to 5$\vec{b}$ in the FiM phase. Good agreement is found between the series of low intensity Raman active modes of the FiM phase measured at $6$~T and the DFT calculations for the fully relaxed 1$\times$5$\times$1 orthorhombic cell with the FiM magnetic order (Fig.~\ref{fig:Struct-and-order}f), as seen in Fig.~\ref{fig:Raman-and-IR}c. The calculated parameters of the relaxed FiM cell are presented in Table S3 of Supporting information.

Increasing further the magnetic field, this series of low intensity features persists up to $19$~T where it abruptly changes to another series, which is strongly reminiscent of the one observed at $0$~T. Comparing the position of the peaks marked with arrows in the Raman scattering response for $0$ and $20$~T in Fig.~\ref{fig:Raman-and-IR}a, all the peaks present at $20$~T appear to be slightly hardened versions of the modes at 0 T (some of the low-intensity features are only visible at 0~T). This series persists up to $27$~T and, then progressively vanishes. From $27$~T up to $30$~T, the Raman scattering response is only composed of the three high-intensity $A_g$ phonon peaks of the FM phase. We observe a blue shift of the three $A_g$ modes at 30~T with respect to those at 20~T. The full magnetic field dependence showing a $B^2$ hardening of their energies is presented in Fig.~S5 of Supporting Information.  The comparison of the experimental Raman scattering response at 30~T to the DFT calculated spectra for the 1$\times$1$\times$1 orthorhombic cell with the FM magnetic order shown in Fig.~\ref{fig:Struct-and-order}i is presented in Fig.~\ref{fig:Raman-and-IR}d. The calculated parameters of the relaxed FM cell are presented in Table S4 of Supporting information.

The additional features also show magnetic hysteresis. As shown in Fig.~\ref{fig:MvsB-and-Raman}d, when sweeping down the magnetic field after having reached the magnetization saturation at $B=30$~T, the magnetic field values at which the series changes are different. From Fig.~\ref{fig:MvsB-and-Raman}c and d we can thus identify four different regions in the magneto-Raman scattering response of bulk CrOCl, which we have labeled with arabic and primed arabic labels, from 1 to 4 and 1' to 4', respectively. The main difference between increasing or decreasing the magnetic field concerns the critical field that defines the transition from region 2 and 3 (19 T), to be compared to the transition from 2' and 3' (15.5~T). The series of additional Raman active modes can hence be used as a tool to evaluate/characterize magnetic hysteresis.

Such additional features are also observed with different selection rules when measuring the infrared (IR) transmission of bulk CrOCl. The low temperature IR transmission spectra measured at $B=0$~T (AFM phase) and at $B=5$~T (FiM phase) are presented in Fig.~\ref{fig:Raman-and-IR}e. They present distinct sharp absorption features which correspond to additional IR active modes, distinct from the Raman active series, and the series abruptly changes to another series at the magnetic phase transition. They offer a deeper insight into the phonon band structure as they provide the IR active modes. The magnetic evolution of the IR response up to B=16~T is presented in Fig.~S6-S7 of the Supporting Information.

The comparison between the seven magnetic field regions defined from the magnetization versus magnetic field measurements (labeled with roman numbers) and the ones defined from the Raman scattering measurements are shown in Fig.~\ref{fig:MvsB-and-Raman}e. From this we discuss the nature of the magnetic phases associated to the regions labeled IV, V, and VI in Fig.~\ref{fig:MvsB-and-Raman}a and e.
The increase in magnetization in the $10-14$~T range (region IV) is certainly a canted phase with the same periodicity (5$\vec{b}$) as the FiM phase. Its magnetic order is represented schematically in Fig.\ref{fig:Struct-and-order}g.
The increase in magnetization in the $20.5-30$~T range (region VI) must be a canted phase with the periodicity (4$\vec{b}$) of the AFM phase. The magnetic order of this canted AFM phase is represented schematically in Fig.\ref{fig:Struct-and-order}h.
The hysteresis cycle in the $14-20.5$~T range (region V) corresponds to a competition between these two canted phases. At this stage, the spin flop phase detected in the magnetization measurements in the $3-3.85$~T range (region II in Fig.~\ref{fig:MvsB-and-Raman}a cannot be distinguished from the AFM 1$\times$4$\times$1 phase as both phases share the same (magnetic) periodicity.

\section{Zone-Folded Phonons}
\label{sec:zfp}

The additional Raman and infrared active modes that appear at low temperature and under a magnetic field can be, in general, attributed to small distortions that change the crystallographic symmetry of the unit cell. In bulk CrOCl, the paramagnetic and the FM phases have the same orthorhombic structure with $Pmmn$ space group. The AFM magnetic order is monoclinic with $P2_1/m$ space group and the FiM phase is orthorhombic with the $Pmmn$ space group. As already mentioned, the AFM unit cell is four times larger along the $\vec{b}$ axis than the paramagnetic one and the FiM unit cell five times larger. One can thus wonder if the additional features appearing in the Raman scattering response could be understood as additional modes that appear at $\Gamma$ as a result of the zone folding of the 1$\times$1$\times$1 cell. Indeed, from group theory analysis, there are nine Raman active phonons in the paramagnetic or FM phase, 36 in the AFM phase, and 45 in the FiM phase.

The phonon band structure of the fully relaxed 1$\times$1$\times$1 FM cell is shown in Fig.~S12 of the Supporting information. The result of folding this band structure in the four and five times smaller Brillouin zones is also shown. They are compared, side-by-side, whith those of the corresponding
fully relaxed AFM and FiM phonon bands. One can observe small energy differences and band anti-crossings which are a consequence of the small structural distortions produced by the magnetoelastic interactions. It is important to note that even if the phonons generated by the strict zone-folded process of the 1$\times$1$\times$1 cell will appear as Raman or IR active by the group theory analysis, their intensity will be zero. No new modes can appear by \textit{unnecessarily} changing the size of the unit cell. It is the small displacements of the atomic positions due to the magnetoelastic interactions, evidenced by the slight change of their energies that make them visible.

\begin{figure*}[ht!]
    \includegraphics[width=1.0\linewidth,angle=0,clip]{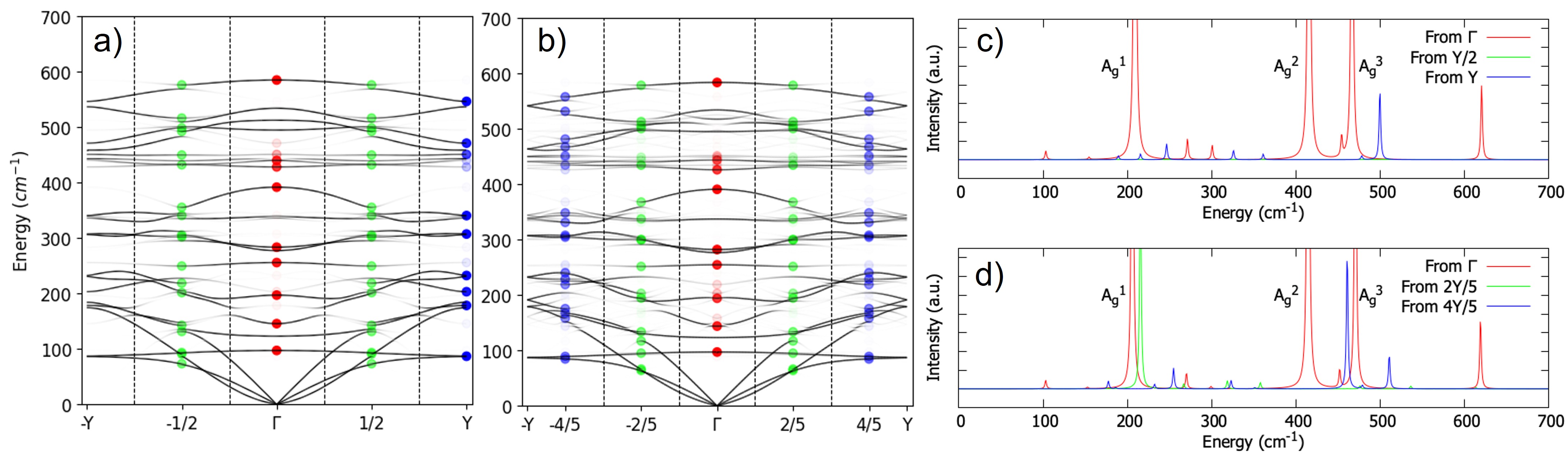}
    \caption{
        a) Calculated phonon band structure of the fully relaxed 1$\times$4$\times$1 AFM phase unfolded in the Brillouin zone of the 1$\times$1$\times$1 unit cell. The gray hue of the lines represents the weight in the unfolding procedure (see text). The dots at $\Gamma$ (red), $\pm$Y/2 (green), and Y (blue) are the 36 Raman active modes of the 1$\times$4$\times$1 AFM phase unfolded in the 1$\times$1$\times$1 Brillouin zone.
        b) Similar unfolding process performed for the fully relaxed 1$\times$5$\times$1 FiM band structure. The dots at $\Gamma$ (red), $\pm$2Y/5 (green), and $\pm$4Y/5 (blue) are the 45 Raman active modes of the 1$\times$5$\times$1 FiM phase unfolded in the 1$\times$1$\times$1 Brillouin zone.
        c) Calculated intensity of the 36 Raman active modes shown in panel a.
        d) Calculated intensity of the 45 Raman active modes shown in panel b.
    \label{fig:Unfolding}}
\end{figure*}

The phonon band structure of the fully relaxed 1$\times$4$\times$1 AFM and 1$\times$5$\times$1 FiM magnetic systems (see Fig.~S12 of the Supporting information) can be \textit{unfolded} in the Brillouin zone of the 1$\times$1$\times$1 cell through the procedure proposed by Allen et al.~\cite{Allen2013}. The $L$ phonon mode with wave-vector $\vec{K}$ in the AFM (FiM) Brillouin zone and energy $\omega(\vec{K}L)$ will appear with a weight
$$
W_{\vec{K}L}(\vec{G}) =
           \frac{1}{\mathcal{N}}\ \sum_{j=1}\sp\mathcal{N}\ \bra{\vec{K}L}{\hat{T}(\vec{r}_j)}\ket{\vec{K}L}\
                                      \text{e}\sp{-i (\vec{K} + \vec{G}) \cdot \vec{r}_j}
$$
and the same energy at wave-vector $\vec{G}$ in the Brillouin zone of the 1$\times$1$\times$1 cell. Here $\mathcal{N}$ is the number of crystal unit cells in the magnetic supercell (4 for AFM and 5 for FiM) and $\hat{T}(\vec{r}_j$) is the translation operator of 1$\times$1$\times$1 unit cells labeled with index $j$ along the $\vec{b}$-axis.
The resulting phonon band structures for the AFM and FiM systems are shown in Fig.~\ref{fig:Unfolding}a and b where the hue of the gray lines represents the weight of the unfolding procedure.
There is a strong resemblance of the unfolded phonon band structures with the band structure of the 1$\times$1$\times$1 cell shown in Fig.~S12 of the Supporting Information.
The 36 red, green, and blue dots in Fig.~\ref{fig:Unfolding}a and 45 in Fig.~\ref{fig:Unfolding}b are the Raman active modes with $\vec{K}=\Gamma$ unfolded in the Brillouin zone of the 1$\times$1$\times$1 cell.
As expected, the ZFP of the AFM cell come from $\Gamma$ (red), $\pm$ Y/2 (green), and Y (blue) in the 1$\times$1$\times$1 cell and the ZFP of the FiM cell come from $\Gamma$ (red), $\pm$2Y/5 (green), and $\pm$4Y/5 (blue).

This unfolding procedure allows us to identify the origin of the additional features that appear in the Raman response, and assign them to the points of the 1$\times$1$\times$1 Brillouin zone commensurate with the magnetic periodicity. The result is shown in Fig.~\ref{fig:Unfolding}c and d, where the calculated Raman spectra are plotted with the same colors as in the unfolded band structure. Additional features in the FiM spectra can be assigned to ZFP from $\vec{G} = \Gamma$, 2Y/5, and 4Y/5.
We can see that the apparent splitting of the $A_g\sp1$ and $A_g\sp3$ recently reported~\cite{Gu2022} for the FiM phase, is actually the result of the appearance of ZFP from 2Y/5 and 4Y/5 respectively.

For the AFM phase only ZFP from $\vec{G} = \Gamma$ and Y appear as new features in the Raman spectra. The intensity of the ZFP from $\vec{G} =$ Y/2 is strictly zero.
This effect has its origin in the fact that ZFP from Y/2 are expected for a 4$\vec{b}$ periodicity, such as in the AFM phase but, as already mentioned, the structural distortion of this phase has only a 2$\vec{b}$ periodicity. This result is confirmed by our calculations of the magnetic order of the AFM phase (see Fig.~S9 of the Supporting information). The spin configurations presented in Fig.~\ref{fig:Struct-and-order}e-i are schematics and correspond to the simplest ones compatible with the size of the magnet unit cells deduced from Raman scattering measurements. There could be additional interactions leading to non collinear spin arrangements such as hidden Dzyaloshinskii–Moriya Interactions~\cite{Yang2022,Cui2024}, which are not taken into account in the present work.

\section{Magneto-cristalline anisotropy}
\label{sec:aniso}

\begin{figure*}[ht!]
    \includegraphics[width=0.95\linewidth,angle=0,clip]{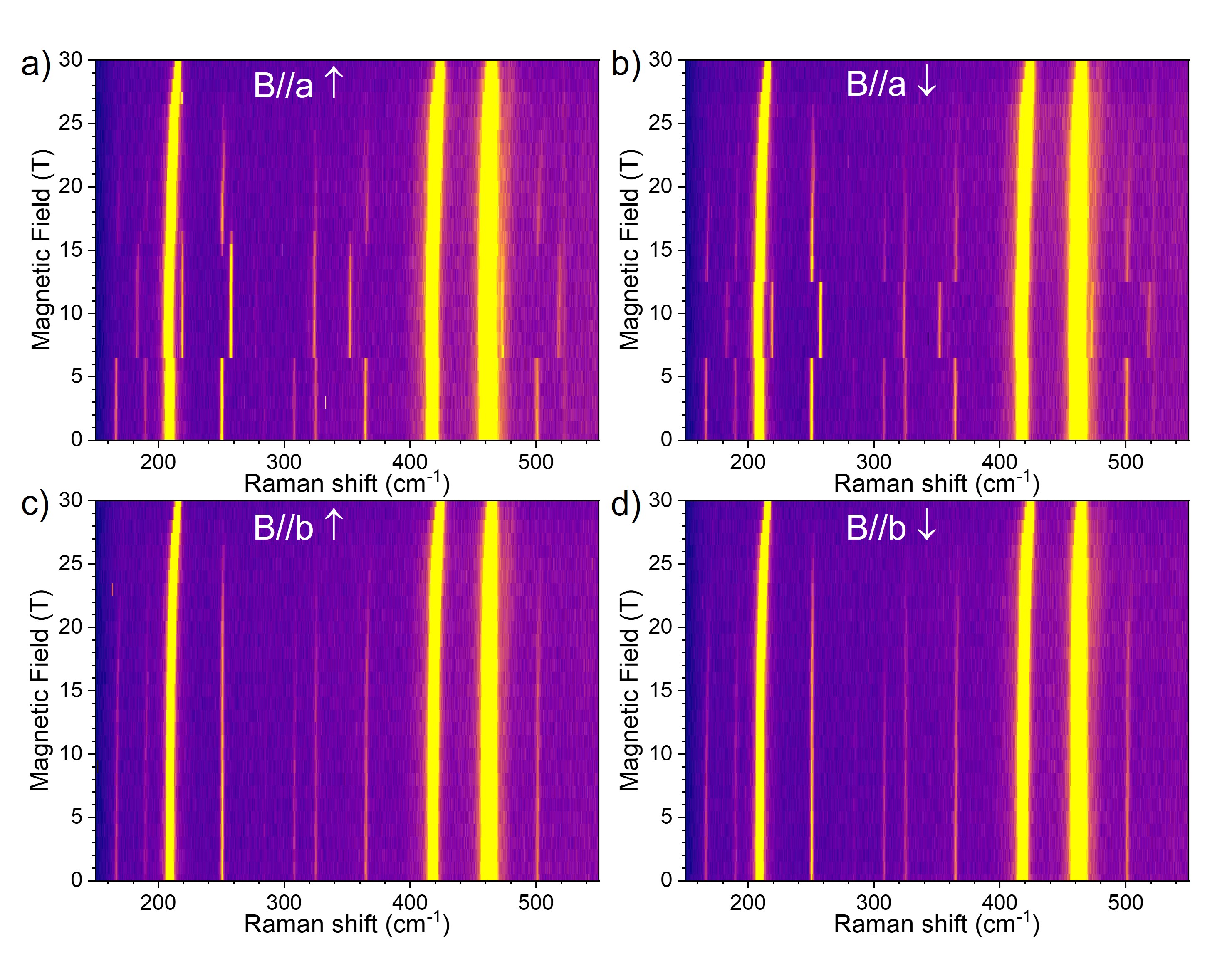}
    \caption{\label{fig:anisotropy} False color map of the low temperature magneto-Raman scattering response of bulk CrOCl with the B-field applied along the long axis of the needles and sweeping from B=0 to 30 T a) and from B=30 T to 0 b), and with the magnetic field applied along the short axis of the needle while sweeping from B=0 to 30 T c) and from B=30 T to 0 d). The spectra are measured with a 1~T field increment/decrement.}
\end{figure*}

Having established previously that one can trace the various magnetic phase transitions using ZFP spectroscopy when the B field is applied along the $\vec{c}$ axis, we now consider a magnetic field along the in-plane $\vec{a}$ and $\vec{b}$ directions. Exfoliation of bulk CrOCl produces lamellar needles with a long ($\vec{a}$ axis) and a short axis ($\vec{b}$ axis), similar to the case of CrSBr~\cite{Wilson2021}. Applying the field along $\vec{a}$, the $B$-sequence of ZFP spectra essentially changes by the values of the critical fields separating the FiM and the canted-FiM phases (regions 1 and 2 or 1' and 2' in the Raman response in Fig.~\ref{fig:MvsB-and-Raman}), and separating the canted-FiM and the canted-AFM phases (regions 2 and 3 or 2' and 3' in Fig.~\ref{fig:MvsB-and-Raman}). The critical fields are 6.5~T and 15.5~T upon increasing field, instead of 3.5~T and  19~T (compare Fig.~\ref{fig:MvsB-and-Raman}c and Fig.~\ref{fig:anisotropy}a). When sweeping the B field from $30$~T down to $0$, a hysteresis is also observed with critical fields observed at $12.5$~T and at $6.5$~T, again different from the case of a field applied along $\vec{c}$ (compare Figs~\ref{fig:MvsB-and-Raman}d and \ref{fig:anisotropy}b). When the field is applied along the short axis of the needle, the observed evolution is distinct, with no change of the ZFP series up to the highest value of the magnetic field, see Fig.~\ref{fig:anisotropy}c,d, suggesting that the FiM phase cannot be stabilized in this configuration. For both in-plane field orientations, the magneto-elastic interaction starts to be effective and shifts the three main phonon modes above $25$~T. These experiments set bulk CrOCl in the class of bi-axial magnetic systems.

\section{Conclusion}

This study establishes CrOCl as a frustrated van der Waals magnet with a rich phase diagram governed by competing magnetic interactions. Using magnetometry, magneto-Raman, and infrared spectroscopy, along with density functional theory, we identified multiple magnetic phases with large unit cells, leading to distinct zone-folded phonon series. These results demonstrate the strong coupling between the non-trivial spin configurations found in bulk CrOCl and lattice vibrations, highlighting the role of magnetoelastic interactions in frustrated magnets. Our findings expand the understanding of 2D magnetism, particularly in materials where frustration leads to unconventional spin states. The observed magnetic hysteresis and field- induced transitions provide a pathway to controlling spin configurations through external stimuli.
Beyond fundamental research, these insights have implications for next- generation technologies. The ability to manipulate spin configurations in van der Waals magnets could have a significant impact on data storage, enabling ultra-dense and energy-efficient memory devices. These results pave the way for future studies on spin-lattice coupling and quantum information processing, positioning frustrated magnets as key materials in advanced spin-based technologies.

\section{Experimental and computational Details}
\label{sec:details}

\subsection{Growth of bulk CrOCl}
CrOCl was prepared by chemical vapor transport (CVT) in quartz glass ampoule using chromium oxide and chromium chloride.
In quartz ampoule ($40\times200$ mm) were placed anhydrous chromium(III) chloride ($99.9\%$, Strem, USA) and chromium oxide ($99.9\%$, Sigma-Aldrich, Czech Republic) in stochiometric ratio corresponding to $20$~g of CrOCl together with $0.3$~g of $HgCl_2$ (99.9\%, Sigma-Aldrich, Czech Republic). The ampoule was melt sealed under high vacuum ($1\times10^{-3}$~Pa) by oxygen-hydrogen welding torch. The ampoule were first pre-reacted in crucible furnace by gradual heating up to $900^\circ\text{C}$. Subsequently, the ampoule was placed in a two zone horizonal furnace. First the source zone was heated at $800^\circ\text{C}$ while the growth zone was heated at $950^\circ\text{C}$. After $4$ days the thermal gradient was reversed and the source zone was heated at $900^\circ\text{C}$ and the growth zone at $850^\circ\text{C}$. Over the period of 15 days the temperature of the source zone were increased on $950^\circ\text{C}$ and the growth zone temperature was decreased to $800^\circ\text{C}$. Finally the ampoule was cooled down to room temperature and opened in an argon filled glovebox.

\subsection{Magnetization measurements}

Pulsed-field magnetization up to $60$~T was measured at the  Dresden High Magnetic field Laboratory. The field was produced by a magnet solenoid energized by a single $1.44$~MJ capacitor module. The raise time of the field pulse was $7$~ms with a total pulse length of $25$~ms.  Magnetization was obtained by integrating the voltage induced in a compensated coil system surrounding the sample. Detailed description of the technique is provided in~[\onlinecite{Skourski2011}].

\subsection{Raman scattering}

The magneto-Raman scattering response of bulk CrOCl was measured using a home made experimental set-up. The excitation laser is focused on the sample by a long working distance objective ($12$~mm) which is also used to collect scattered signals which are sent using free beam optics to a grating spectrometer equipped with a liquid nitrogen cooled CCD. Three volume Bragg filters are used in series before the spectrometer to remove the elastic scattering. This set-up is then placed in a metallic tube filled with helium exchange gas and inserted in liquid helium. We use an optical power of $50$~$\mu$W focused on $\sim 1$~$\mu$m and typical acquisition time for the Raman scattering response of $300$~sec.

\subsection{Infrared measurements}

The infrared response has been measured using a Fourier transform spectrometer coupled to the resistive magnet using an oversized waveguide. The sample is placed in a closed tube filled with helium exchange gas and the signal is collected by a composite silicon bolometer placed behind the sample. We use a rotating sample holder to change in-situ between the sample and a reference to measure the absolute transmission of the sample. Static magnetic fields up to $B=16$~T have been produced by a superconducting solenoid.

\subsection{Magneto-optical Kerr effect mesurements}

Visible magneto-optical Kerr spectroscopy at $4$~K was measured upon reflection in polar configuration with magnetic field applied perpendicular to the sample surface and nearly normal light incidence. We employed custom-built magneto-optical spectrometer based on rotating analyzer technique coupled to the Quantum Design Physical Property Measurements System. The applied magnetic field ranged from $-6$ to $6$~T and the measured spectral region covered energies from $1.2$ to $4$ eV.

\subsection{Computational details}

Phonon calculations were performed using the finite-displacements in supercell method as implemented in the Phonopy package \cite{phonopy-phono3py-JPCM,phonopy-phono3py-JPSJ}.
The interatomic forces were extracted from density-functional theory (DFT) calculations using Quantum Espresso \cite{giannozzi_quantum_2009,giannozzi_advanced_2017}. The Perdew-Burke-Ernzerhof (PBE) \cite{perdew_generalized_1996} version of generalized gradient approximation (GGA) \cite{perdew_generalized_1996-1} for the exchange correlation functional has been used with a plane wave basis set and norm-conserving pseudopotentials \cite{hamann_norm-conserving_1979,hamann_optimized_2013}. In this way, scalar-relativistic spin-polarized calculations are performed for every magnetic cell. Plane wave kinetic energy cutoffs of $150$~Ry and $600$~Ry were used for wave functions and the charge density respectively. Because only the $\Gamma$-point of the Brillouin zone is of interest in optically excited phonons, and since $\Gamma$-point calculations are always "exact" \cite{parlinski_first-principles_1997}, the magnetic primitive cell size was used as the smallest supercell size possible for lighter calculations.

Calculations have been performed for three different magnetic structures. The FM phase (see Fig.~\ref{fig:Struct-and-order}i) was calculated in the $Pmmm$ cell (space group $59$) with the room temperature experimental \cite{Christensen1975} lattice constants $a$ = 3.863 \AA,  $b$ = 3.863 \AA, and $c$ = 7.694 \AA. In this phase, the Wyckoff positions are $2a$, $2b$, and $2b$ for the Cr, O, and Cl atoms respectively. There are a total of six atoms in the unit cell. The AFM phase (see Fig.\ref{fig:Struct-and-order}e) was calculated in a 24 atoms 1 $\times$ 4 $\times$ 1 cell and the FiM phase (see Fig.\ref{fig:Struct-and-order}f) in a 30 atoms 1 $\times$ 5 $\times$ 1 cell. The AFM cell has $P21/m$ (space group $11$) symmetry and the FiM cell $Pmmm$ (space group $59$) symmetry.

Using the Monkhorst-Pack method \cite{monkhorst_special_1976}, the Brillouin zones corresponding to the FM, AFM and FiM cells were sampled with 9 $\times$ 11 $\times$ 5, 9 $\times$ 3 $\times$ 5, and 9 $\times$ 2 $\times$ 5 k-point meshes respectively, in accordance with cell size. Shifted k-meshes were used for relaxations only, in order to reduce by symmetry the number of treated k-points. The non-shifted meshes containing the $\Gamma$-point were used for force calculations.
Raman intensity calculations have been performed using the Phonopy Spectroscopy module \cite{skelton_lattice_2017}. This method recasts the Raman intensity tensors in terms of the macroscopic high-frequency dielectric tensor using the central-difference scheme. Based on the equations of \cite{skelton_lattice_2017}, it was found that the minimal precision to properly treat low-intensity modes required an accuracy on the computed static dielectric tensors of $10^{-4}$\ au. Non-analytical term corrections have been included in the phonon band structure calculations using the dipole-dipole interactions formalism \cite{Pick1970,Gonze1994,Gonze1997}.

\acknowledgments
This work was partially supported by LNCMI and HLD-HZDR, members of the European Magnetic Field Laboratory (EMFL). This work is also partially supported by France 2030 government investment plan, managed by the French National Research Agency under Grant Reference PEPR SPIN–SPINMAT ANR-22-EXSP-0007, by ANR-23-QUAC-0004 and by CEFIPRA CSRP Project No. 7104-2. Z.S. and B.W. were supported by ERC-CZ program (project LL2101) from Ministry of Education Youth and Sports (MEYS) and by the project Advanced Functional Nanorobots (reg. No. CZ.$02.1.01/0.0/0.0/15-003/0000444$ financed by the ERDF).

\end{document}